\documentclass[aps,prl,showpacs,preprintnumbers,amssymb]{revtex4}

\usepackage{graphicx}
\usepackage{dcolumn}
\usepackage{bm}

\begin{document}

\title{Classical self-forces in a space with a dispiration}
\author{V. A. De Lorenci and E. S. Moreira Jr.}
\affiliation{ Instituto de Ci\^encias -
Escola Federal de Engenharia de Itajub\'a \\
Av. BPS 1303 Pinheirinho, 37500-903 Itajub\'a, MG -- Brazil \\
(Email addresses: {\tt lorenci@efei.br, moreira@efei.br})}

\date{\today}

\begin{abstract}
We derive the gravitational and electrostatic self-energies
of a particle at rest in the background of a cosmic
dispiration (topological defect), finding that the particle may 
experience potential steps, well potentials or potential barriers depending
on the nature of the interaction and also on certain properties of the
defect.  The results may turn out to be useful in 
cosmology and condensed matter physics.
\end{abstract}
\maketitle

\section{Introduction}

Spontaneous symmetry breaking 
could have caused the appearance of topological defects in the very early
universe \cite{Kibble}. Spacetimes corresponding to 
defects such as cosmic strings, domain walls and monopoles represent
solutions of the Einstein equations, giving rise to
many gravitational and cosmological effects where particle production during
phase transitions (when the formation of defects takes place)  \cite{Parker},
gravitational lensing \cite{Vilenkin} and galaxy formation  are examples
\cite{Zel'dovich,Vilenkin2,Vilenkin3}.

The formalism of general relativity to deal with spacetime defects
has been used 
in the context of condensed matter physics as a 
tool to tackle certain problems involving defects in solids
\cite{Kroner,Katanaev,Balachandran,Lazar}.
(Ref. \cite{Lazar}, for example, has examined analogous aspects shared
by a screw dislocation in solids, a cosmic string and a
magnetic vortex, discussing also the dislocation
theory in solids as a three-dimensional theory of gravity.)
Following this program quantum field
theory \cite{Bausch} and 
classical field theory (see Ref. \cite{Carvalho} and references
therein) in the bulk of solids with linear defects 
(such as disclinations and dislocations)
have recently been considered. 

This work examines classical self-force effects on a test particle at rest
in a spacetime with a cosmic dispiration (a screw dislocation plus a
disclination \cite{Letelierx,Tod,Moraesx}), which is locally flat
[cf. Eq. (\ref{mmet}) below]. Self-forces are
more surprising in a locally flat background
\cite{Linet86,Smith,Gal'tsov,Khusnutdinov,khu01} 
rather than in one with non
vanishing local curvature, such as a black hole \cite{Burko,Linet,Poisson}.
In a spacetime with a dispiration, self forces arise due to
the non trivial global geometry which distorts the
test particle gravitational and electrostatic fields.

In the next section the Green function 
for the Poisson equation in a background with a dispiration is
obtained, and used in the following section to develop
an analytical and numerical study of the gravitational and electrostatic
self-forces on a test particle at rest. A discussion of rather curious
self-force effects is presented in the last section.

\section{The Green function}

The geometry of a cosmic dispiration is characterized 
by the Minkowski line element written in cylindrical
coordinates, 
\begin{equation}
ds^{2}=dt^{2}-dr^2 -r^2 d\varphi^2 - dZ^2,
\label{mmet}
\end{equation}
and by the non trivial identification
$(t,r,\varphi,Z) \sim (t,r,\varphi+2\pi\alpha,Z+2\pi\kappa)$,
where $\alpha>0$ and $\kappa\geq 0$ are 
parameters corresponding to
a disclination and a screw dislocation, respectively.
By defining new space coordinates
$\theta := \varphi / \alpha$ and $z := Z - (\kappa/\alpha)\varphi$, 
the line element of the space section becomes \cite{Letelierx,Tod}
\begin{equation}
dl^2=dr^2 + \alpha^2 r^2 d\theta^2 + (dz + \kappa d\theta)^2,
\label{6}
\end{equation}
and the usual identification
$(r,\theta,z) \sim (r,\theta +2\pi,z)$ must be observed.
Clearly, when $\alpha=1$ and $\kappa=0$ the space 
becomes Euclidean. 

Following Ref. \cite{Smith},
in order to obtain self-energies in the space 
corresponding to Eq. (\ref{6}), one evaluates  
the Green function $G(x,x')$, which is solution of
\begin{equation}
\bigtriangledown^2 G(x,x') = -\frac{4\pi}{\sqrt{g}} \delta(r-r')
\delta(\theta - \theta')\delta(z -z'),
\label{32}
\end{equation}
with 
\begin{displaymath}
\bigtriangledown^2=
\frac{1}{r}\frac{\partial}{\partial r}\left(r
\frac{\partial}{\partial r}\right) +
\frac{1}{\alpha^2 r^2}\frac{\partial^2}{\partial \theta^2}
+ \left(1+\frac{\kappa^2}{\alpha^2 r^2}\right)\frac{\partial^2}{\partial z^2}
-\frac{2\kappa}{\alpha^2 r^2}
\frac{\partial^2}{\partial\theta\partial z},
\end{displaymath}
and $g=\alpha^{2} r^{2}$.
The regular eigenfunctions of the operator $-\bigtriangledown^2$
are given by  
\begin{equation}
\phi_{n,\nu,\mu}(x) = \frac{1}{2\pi\sqrt{\alpha}} 
J_{|M|}(\mu r)e^{-i n \theta}e^{i \nu z},
\label{22}
\end{equation}
where
$n$ is an integer, $\nu$ is a real number,
\begin{equation}
M:=\frac{n+\nu\kappa}{\alpha}, 
\label{mfactor}
\end{equation}
$\mu$ is a positive real number and 
$J_{\beta}$ denotes the Bessel function of the first kind.
The corresponding eigenvalues are 
$\lambda_{\mu,\nu}=\mu^{2}+\nu^{2}$.
By considering the completeness relation of the 
eigenfunctions $\phi_{n,\nu,\mu}(x)$,
it follows that $G(x,x')$ can be expressed as \cite{hel86,pon}
\begin{equation}
G(x,x') = 4\pi i\int_0^\infty dt \sum _{n=-\infty}^{\infty}
\int_{-\infty}^{\infty} d\nu \int_{0}^{\infty}d\mu\, \mu\
e^{-i t \lambda_{\mu,\nu}}\ \phi_{n,\nu,\mu}(x)\ \phi^{*}_{n,\nu,\mu}(x').
\label{transp1}
\end{equation}
In order to extract from Eq. (\ref{transp1}) 
self-force effects one needs
to consider $G(x,x)$ \cite{Smith}. 
Accordingly, by evaluating the integral over
$\mu$ \cite{Gradshteyn} and defining $T:=it$, Eq. (\ref{transp1}) yields  
\begin{equation}
G(r) = \frac{1}{2\alpha \pi}\int_{-\infty}^\infty
\!d\nu \int_{0}^{\infty} \!\frac{dT}{T}e^{-T\nu^2-r^2/2T}
\sum _{n=-\infty}^{\infty} I_{|M|}(r^2/2T),
\label{42}
\end{equation}
where $I_{\beta}$ denotes the modified Bessel function of
the first kind. Using a convenient integral representation 
for $I_{\beta}$ \cite{Gradshteyn} it follows that 
\begin{equation}
\sum_{n=-\infty}^{\infty} I_{|M|}(r^2/2T)
= \alpha e^{r^2/2T} -\frac{1}{\pi} \int_{0}^{\infty} \!dx\;  
e^{-(r^2/2T)\cosh x} \sum_{n=-\infty}^{\infty} \sin (|M|\pi)\
e^{-|M|x},
\label{67}
\end{equation}
which holds for $\alpha>1/2$
[smaller values of $\alpha$ can be considered by taking into account
terms that were omitted in Eq. (\ref{67})],
resulting 
\begin{eqnarray}
G(r) &=& \frac{1}{2\pi}\int_{0}^{\infty} \!\frac{dT}{T} 
\int_{-\infty}^\infty
\!d\nu\ e^{-T\nu^2}
\nonumber \\
&&-\frac{1}{2\alpha\pi^2}
\int_{0}^{\infty} \!\frac{dT}{T}\ e^{-r^2/2T}
\sum_{n=-\infty}^{\infty} \int_{-\infty}^{\infty} d\nu\  e^{-T\nu^2}
\sin (|M|\pi)\int_{0}^{\infty}dx\
e^{-(r^2/2T)\cosh x -|M|x}.
\label{69}
\end{eqnarray}
The first term on the right hand side (r.h.s.) of  Eq. (\ref{69}) is an
infinite constant and does not affect the evaluation of self force effects
(cf. next section). One therefore simply drops this term,
which amounts to renormalize Eq. (\ref{69})
with respect to the Euclidean contribution
[by considering Eq. (\ref{mfactor}) 
it should be noticed that the second term on
the r.h.s. of Eq. (\ref{69}) vanishes when $\alpha=1$ and $\kappa=0$].
By inserting
$\delta(\lambda-M)$
in Eq. (\ref{69}) and using Poisson's formula,
one is able to perform the integration over $\nu$ \cite{Gradshteyn}.
Finally, the integrations over $\lambda$ and $T$ 
are evaluated \cite{Gradshteyn}, resulting in
\begin{eqnarray}
G_{\scriptscriptstyle ren}(r)&=& -\frac{\ln(2)}{\pi r}
-\frac{2}{r}\sum_{n=1}^{\infty}
\int^{\infty}_0 dx 
\frac{x^2-\pi^2(4\alpha^2 n^2-1)}{\left[\pi^2(2\alpha n+ 1)^2
+ x^2\right] \left[\pi^2(2\alpha n - 1)^2 +x^2\right]
\left[\cosh^2(x/2) + (n\pi\kappa/r)^{2}\right]^{1/2}}.
\label{95}
\end{eqnarray}

\section{The self-forces}

Self-force effects are better appreciated when described in terms
of the flat coordinates appearing in Eq. (\ref{mmet}).
The gravitational and electrostatic  self-energies of a point 
particle of mass $m$ and charge $q$ can be expressed in terms of 
$G_{\scriptscriptstyle ren}(r)$ as 
\begin{equation}
U_{m}(r) = -\frac{{\cal G} m^2}{2}G_{\scriptscriptstyle ren}(r)
\label{98}
\end{equation}
and
\begin{equation}
U_{q}(r) = \frac{q^2}{2}G_{\scriptscriptstyle ren}(r),
\label{98b}
\end{equation}
respectively. These expressions arise from the corresponding
Poisson equations by taking the test particle as source
\cite{Smith,Jackson}.
The self-forces can  then be obtained by taking
minus the gradient of the self-energies. However, usually the 
behavior of a test particle under the action of a force
are better understood by considering the corresponding  potential 
energy, rather than the force itself.
Therefore the analysis below will be based upon Eqs. (\ref{95}),  
(\ref{98}) and (\ref{98b}).

As $\kappa/r\rightarrow 0$, the summation in Eq. (\ref{95})
can be evaluated by considering the power series expansion
of $\psi(z)$ (the logarithmic derivative of the gamma function)
and its properties, yielding as leading contribution
\begin{equation}
G_{\scriptscriptstyle ren}(r)=
-\frac{1}{2\pi\alpha\, r}\sin(\pi/\alpha)
\int^{\infty}_0 dx\ 
\frac{1}{\cosh(x/2)[\cosh(x/\alpha)-\cos(\pi/\alpha)]}.
\label{ap1}
\end{equation}
As $\kappa/r\rightarrow \infty$, on the other hand,  the leading 
contribution is now given by 
\begin{equation}
G_{\scriptscriptstyle ren}(r)=-\frac{\ln(2)}{\pi r}.
\label{ap2}
\end{equation}

Apart from these asymptotic cases,
the dependence of  $G_{\scriptscriptstyle ren}(r)$ on $r$ is non trivially
hidden in Eq. (\ref{95}) and a numerical analysis is required.
The plots (where units were omitted) show how $-G_{\scriptscriptstyle
ren}(r)$ varies with $r$ for various combinations of values of $\alpha$ 
and $b:=2\pi\kappa$.
\begin{figure}[htb]
\leavevmode
\centering
\includegraphics{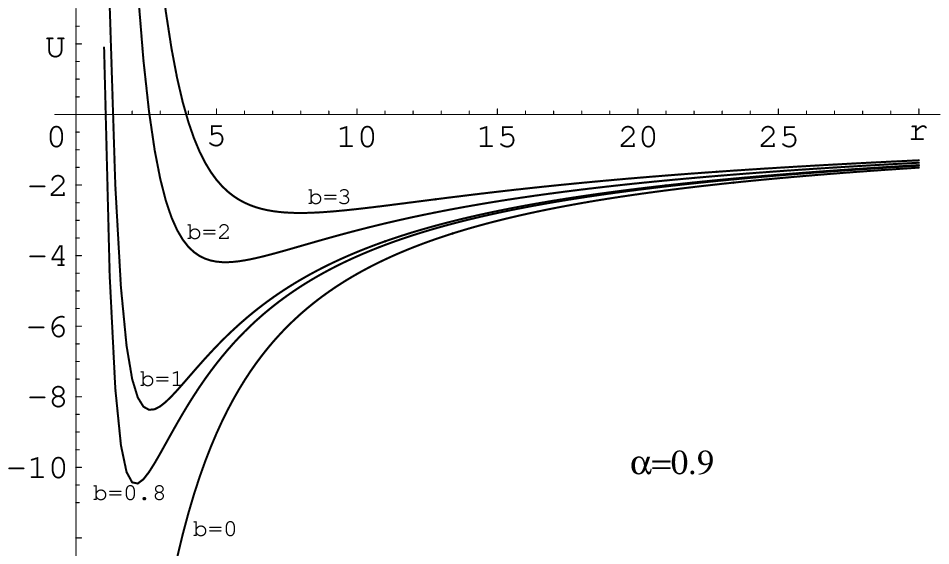}%
\caption{${\rm U} := - G_{\scriptscriptstyle ren}(r) 
\times 10^{-3}$.}
\label{conjunto4}
\end{figure}
\begin{figure}[htb]
\leavevmode
\centering
\includegraphics{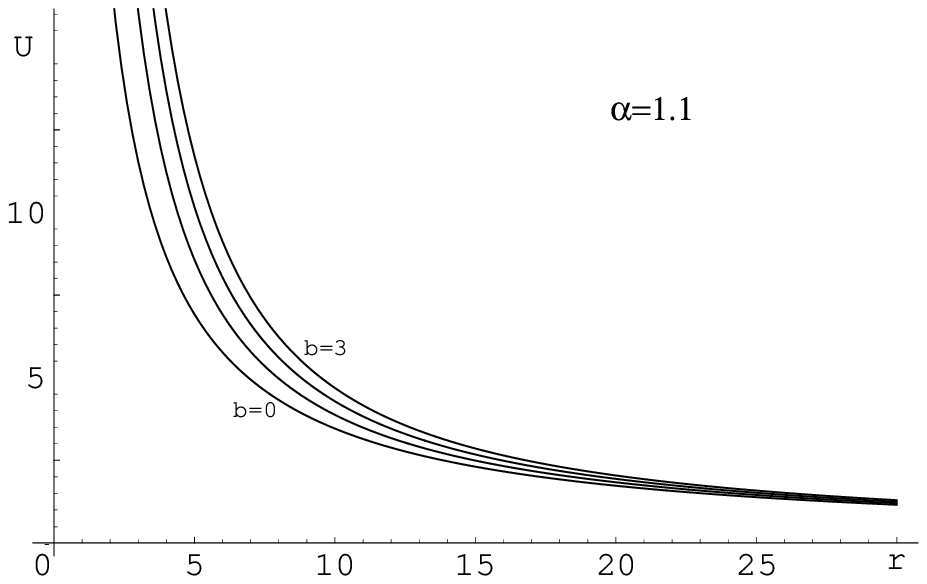}%
\caption{$ {\rm U} := - G_{\scriptscriptstyle ren}(r) 
\times 10^{-3}$. The intermediate plots, from the bottom, correspond
to $b=1$ and $b=2$, respectively.}
\label{conjunto3}
\end{figure}
\begin{figure}[htb]
\leavevmode
\centering
\includegraphics{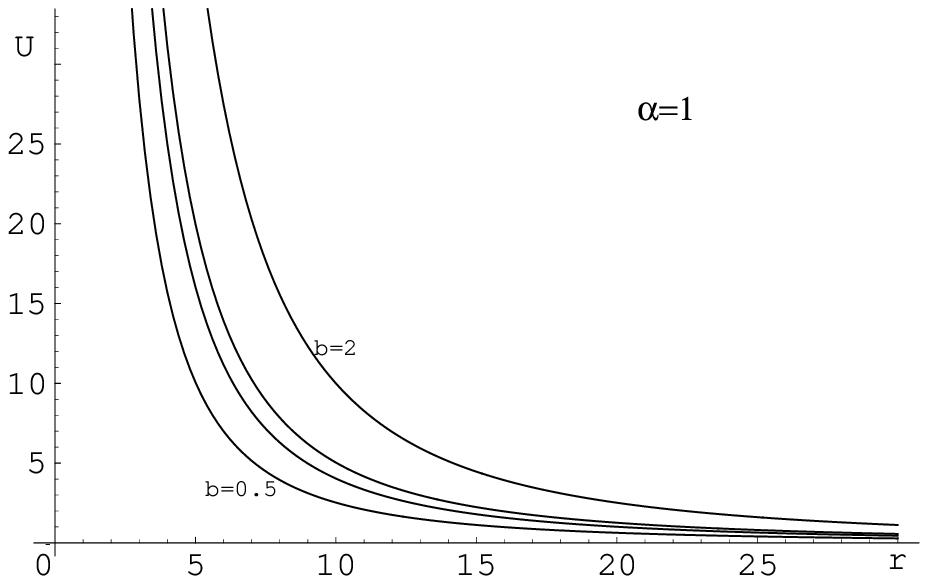}%
\caption{$ {\rm U} := - G_{\scriptscriptstyle ren}(r) 
\times 10^{-4}$. The intermediate plots, from the bottom, correspond
to $b=0.8$ and $b=1$, respectively.} 
\label{conjunto2}
\end{figure}
\begin{figure}[htb]
\leavevmode
\centering
\includegraphics{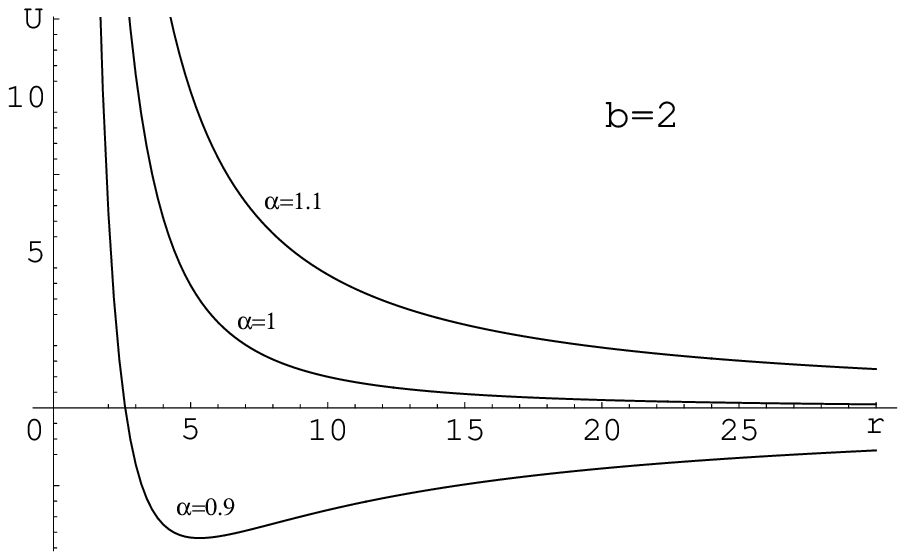}%
\caption{${\rm U} := - G_{\scriptscriptstyle ren}(r) 
\times 10^{-3}$.}
\label{conjunto5}
\end{figure}

\section{Discussion}

By inserting Eq. (\ref{ap1}) in 
Eqs. (\ref{98}) and (\ref{98b}), one
reproduces the known results in the literature 
\cite{Linet86,Smith}.
Namely, the gravitational self-force due to a disclination alone,
i.e. when $\kappa=0$,
is attractive (repulsive) for $\alpha<1$ ($\alpha>1$).
The opposite holds for the electrostatic self-force
(cf. Figs. \ref{conjunto4} and \ref{conjunto3}).

For $\kappa\neq 0$, Eq. (\ref{ap1})  shows that, 
as $r$ gets very large, disclination effects become dominant
(cf. Figs. \ref{conjunto4} and \ref{conjunto3}).
As  $r$ gets very small though, Eq. (\ref{ap2})
shows that screw dislocation effects rule in a peculiar
way, rendering the  self-forces due to a dispiration 
independent of $\kappa$ and $\alpha$, which are the parameters
characterizing the defect (similar effects are rather familiar in
the context of vacuum polarization \cite{ful89}).
By observing Eqs. (\ref{98}) and (\ref{ap2}),
one sees that the dispiration (more precisely, the screw dislocation)
induces a gravitational barrier, keeping the test particle away from
the defect. From Eqs. (\ref{98b}) and (\ref{ap2})
one sees that the reverse is true for the electrostatic self-force,
i.e., a charged test particle is electrostatically 
attracted by the defect. 

For arbitrary values of the radial coordinate $r$,
numerical and analytical considerations 
show that when $\alpha\geq 1$ (including therefore the
screw dislocation alone for which $\alpha = 1$)
the gravitational self-force induced by the dispiration
is repulsive, whereas the electrostatic self-force is attractive
(cf. Figs. \ref{conjunto3} and \ref{conjunto2}).
When $\alpha<1$, the repulsive gravitational self-force due to
the screw dislocation and the attractive gravitational self-force
due to the disclination dispute, giving rise to an well potential
(a potential step, in the case of the electrostatic self-force).
Quick dimensional considerations reveal that the minimum of
$-G_{\scriptscriptstyle ren}(r)$
is proportional to $-1/\kappa$ whose corresponding $r$
is proportional to $\kappa$ (cf. Figs. \ref{conjunto4} 
and \ref{conjunto5}). It should be noticed that when $\alpha<1$
the behaviors of the self-forces for $\kappa=0$ and
$\kappa\neq 0$ differ radically from each other
as $r\rightarrow 0$
[cf. Eq. (\ref{ap1}), Eq. (\ref{ap2}) and Fig. \ref{conjunto4}].

Before closing, it should be pointed out that Eq. (\ref{ap1})
is the leading contribution as $r\rightarrow \infty$ only when
$\alpha\neq 1$. When $\alpha=1$, Eq. (\ref{ap1}) vanishes
and the sub-leading contribution, which is due to the screw dislocation only,
takes over. By dimensional considerations, one may be tempted to
assume that such a contribution is proportional to $\kappa^2/r^{3}$.
However an  analysis seems to show that 
$-G_{\scriptscriptstyle ren}(r)$ falls much quicker than that 
as $r\rightarrow \infty$ (cf. Fig. \ref{conjunto5}). 
(The representation of  $G_{\scriptscriptstyle ren}(r)$ in
Eq. (\ref{95}) is not very handy to determine the sub-leading
contribution, and an alternative representation may show to be more
useful on this matter.)

It remains to be seen if the results outlined above
have applications in the realms 
of cosmology and condensed matter physics.

\begin{acknowledgments}
The authors are grateful to George Matsas and Fernando Moraes for
valuable discussions. This work was partially supported by {\em Conselho
Nacional de Desenvolvimento Cient\'{\i}fico e Tecnol\'ogico} (CNPq)
of Brazil.
\end{acknowledgments}

\end{document}